\documentclass
[twocolumn,prl,unsortedaddress,superscriptaddress,showpacs]{revtex4}%
\usepackage{amsmath}
\usepackage{amsfonts}
\usepackage{amssymb}
\usepackage{graphicx}
\usepackage[utf8]{inputenc}%
\providecommand{\U}[1]{\protect\rule{.1in}{.1in}}
\providecommand{\U}[1]{\protect\rule{.1in}{.1in}}

\begin{document}
\title{Ghost Free Mimetic Massive Gravity}
\author{Ali H. Chamseddine}
\affiliation{Physics Department, American University of Beirut, Lebanon}
\affiliation{I.H.E.S. F-91440 Bures-sur-Yvette, France}
\email{chams@aub.edu.lb}
\author{Viatcheslav Mukhanov}
\affiliation{Theoretical Physics, Ludwig Maxmillians University,Theresienstr. 37, 80333
Munich, Germany }
\affiliation{MPI for Physics, Foehringer Ring, 6, 80850, Munich, Germany}
\affiliation{LPT de l'Ecole Normale Superieure, 24 rue Lhomond, 75231 Paris cedex, France}
\email{viatcheslav.mukhanov@lmu.de}

\begin{abstract}
The mass of the graviton can be generated using a Brout-Englert-Higgs
mechanism with four scalar fields. We show that when one of these fields is
costrained as in mimetic gravity, the massive gravity obtained is ghost free
and consistent. The mass term is not of the Fierz-Pauli type. There are only
five degrees of freedom and the sixth degree of freedom associated with the
Boulware-Deser ghost is constrained and replaced by mimetic matter to all
orders. The van Dam-Veltman-Zakharov discontinuity is also absent.

\end{abstract}

\maketitle

%\end{titlepage}

In \cite{CMmim} we have proposed an alternative formulation of gravity
explicitly isolating the scale factor in the physical metric $g_{\mu\nu}$ by
rewriting it as
\begin{equation}
g_{\mu\nu}=\tilde{g}_{\mu\nu}\left(  \tilde{g}^{\kappa\lambda}\partial
_{\kappa}\phi\partial_{\lambda}\phi\right)  , \label{1}%
\end{equation}
where $\tilde{g}_{\mu\nu}$ is some auxiliary metric and $\phi$ is a scalar
field. It follows from (\ref{1}) that $\phi$ must obey the constraint%
\begin{equation}
g^{\mu\nu}\partial_{\mu}\phi\partial_{\nu}\phi=1. \label{2}%
\end{equation}
Considering $\tilde{g}_{\mu\nu}$ as a fundamental variable in the
Einstein-Hilbert action we find that Einstein equations are modified to
\begin{equation}
G_{\mu\nu}-T_{\mu\nu}-(G-T)\partial_{\mu}\phi\partial_{\nu}\phi=0, \label{3}%
\end{equation}
where $G_{\mu\nu}$ and $T_{\mu\nu}$ are, respectively, the Einstein tensor and
the energy-momentum tensor for matter. We use the units in which $8\pi G=1.$
The trace part of equation (\ref{3}) vanishes identically as a consequence of
scale invariance $\tilde{g}_{\mu\nu}\rightarrow\Omega_{\mu\nu}^{2}\tilde{g}$
for the auxiliary metric $\tilde{g}_{\mu\nu}$ in equation (\ref{2}). As a
result the modified Einstein equations (\ref{2}) and (\ref{3}) have an extra
solution even in the absence of matter which could mimic cold dark matter in
our Universe. As it was noticed in \cite{Golovnev} this mimetic gravity is
fully equivalent to Einstein gravity with the extra constraint (\ref{2})
implemented in the action using a Lagrange multiplier. The field $\phi$ can be
taken in the synchronous coordinate system, solving constraint (\ref{2}), as
time coordinate. It becomes dynamical only when combined with the longitudinal
mode of gravity. The proposed model does not only give a plausible explanation
for the origin of dark matter but also provides a new approach to resolve
singularities in General Relativity \cite{CMsing}. Moreover, the appearance of
the constrained field $\phi$ has found a justification in the noncommutative
approach to the quanta of geometry \cite{CCM}.

In this letter we will consider the theory of mimetic massive gravity and show
that this theory is ghost free to all orders and describes the massive
graviton with five degrees of freedom which are completely decoupled from
mimetic matter in the linear approximation.

The simplest way of giving mass to the graviton without explicitly spoiling
diffeomorphism invariance, reflecting the freedom in the choice of coordinate
system, is by employing the Brout-Englert-Higgs mechanism with four scalar
fields $\phi^{A},$ $A=0,1,2,3$ \cite{tH,CM}. In Minkowski space-time the
broken symmetry phase,
\begin{equation}
\left\langle \phi^{A}\right\rangle =\delta_{\mu}^{A}x^{\mu}=x^{A}, \label{4}%
\end{equation}
is degenerate with respect to a vacuum choice up to Poincare transformations.
For small perturbations of the fields%
\begin{equation}
\phi^{A}=x^{A}+\chi^{A}, \label{5}%
\end{equation}
the three scalars $\chi^{i}$ are absorbed to give mass to the graviton, while
the fourth field $\chi^{0}$ leads to a ghost, unless the mass term in the
Lagrangian is taken to be of the Fierz-Pauli form \cite{FP} where this ghost
is not excited at the linear level. However, generically, it reappears as a
nonlinear Boulware-Deser ghost on non-trivial backgrounds \cite{BD}. The idea
we put forward in this letter, is to use as one of the four fields, needed for
providing mass to the graviton, the mimetic field $\phi\equiv\phi^{0}.$
Because this field is constrained to be always in the broken symmetry phase
the dangerous degree of freedom is thus replaced by dark matter and the ghost
is avoided to any order in perturbation theory. As we show below the mass term
in this case must be necessarily taken to be different from Fierz-Pauli type.

In massive gravity a central role is played by the induced \textit{metric
perturbations}%
\begin{equation}
\bar{h}^{AB}=g^{\mu\nu}\partial_{\mu}\phi^{A}\partial_{\nu}\phi^{B}-\eta^{AB},
\label{6}%
\end{equation}
the components of which are scalars under diffeomorphism transformations. We
will raise and lower capital indices with the help of the auxiliary
\textit{Minkowski metric}\ $\eta^{AB}=(1,-1,-1,-1),$ so that, for example,
$\bar{h}\equiv\bar{h}_{A}^{A}=$ $\eta_{AB}\bar{h}^{AB}.$

We consider the theory with action%
\begin{align}
I  &  =%
%TCIMACRO{\dint }%
%BeginExpansion
{\displaystyle\int}
%EndExpansion
d^{4}x\sqrt{g}\left(  -\frac{1}{2}R+\frac{m^{2}}{8}\left(  \frac{1}{2}\bar
{h}^{2}-\bar{h}^{AB}\bar{h}_{AB}\right)  \right. \nonumber\\
&  \left.  +\lambda\left(  g^{\mu\nu}\partial_{\mu}\phi^{0}\partial_{\nu}%
\phi^{0}-1\right)  \right)  , \label{7}%
\end{align}
where the last term accounts for the mimetic origin of $\phi^{0}$ and the mass
term has relative coefficient $\frac{1}{2}$ between $\bar{h}^{2}$ and $\bar
{h}^{AB}\bar{h}_{AB}$ in distinction from Fierz-Pauli term where this
coefficient is $1$. The reason for this choice will become clear later. The
equations of motion are obtained, first by varying with respect to $\delta
g^{\mu\nu}$:%
\begin{align}
G_{\mu\nu}  &  =-\frac{m^{2}}{8}\left(  \frac{1}{2}\bar{h}^{2}-\bar{h}%
^{AB}\bar{h}_{AB}\right)  g_{\mu\nu}+\lambda\left(  2\partial_{\mu}\phi
^{0}\partial_{\nu}\phi^{0}\right) \nonumber\\
&  +\frac{m^{2}}{2}\left(  \frac{1}{2}\bar{h}\partial_{\mu}\phi_{A}%
\partial_{\nu}\phi^{A}-\bar{h}_{AB}\partial_{\mu}\phi^{A}\partial_{\nu}%
\phi^{B}\right)  , \label{8}%
\end{align}
and next with respect to $\delta\phi^{A}$:%
\begin{equation}
\nabla^{\mu}\left(  m^{2}\left(  \frac{1}{2}\bar{h}\partial_{\mu}\phi_{A}%
-\bar{h}_{AB}\partial_{\mu}\phi^{B}\right)  +4\lambda\delta_{0A}\partial_{\mu
}\phi^{0}\right)  =0. \label{9}%
\end{equation}
Varying with respect to $\delta\lambda$ gives:%
\begin{equation}
\bar{h}^{00}=0. \label{10}%
\end{equation}
Let us consider small perturbations around Minkowski background, $g_{\mu\nu
}=\eta_{\mu\nu}+h_{\mu\nu}$ and $\phi^{A}=x^{A}+\chi^{A},$ and linearize the
above equations in $h_{\mu\nu}$ and $\chi^{A}$ keeping in mind that $\lambda$
is of first order in perturbations$.$ In equation (\ref{10}) we first set
$A=0$ to get%
\begin{equation}
\partial_{0}\lambda-\frac{m^{2}}{4}\left(  \partial^{\rho}\bar{h}_{\rho
0}-\frac{1}{2}\partial_{0}\bar{h}\right)  =0, \label{11}%
\end{equation}
and then $A=k$ to obtain
\begin{equation}
m^{2}\left(  \partial^{\rho}\bar{h}_{\rho k}-\frac{1}{2}\partial_{k}\bar
{h}\right)  =0. \label{12}%
\end{equation}
The linearized Einstein tensor in $h_{\mu\nu}$ is equal to%
\begin{align}
G_{\mu\nu}\left(  h_{\rho\sigma}\right)   &  =-\frac{1}{2}\left(  \partial
^{2}h_{\mu\nu}-\partial_{\mu}\partial^{\rho}h_{\rho\nu}-\partial_{\nu}%
\partial^{\rho}h_{\rho\mu}+\partial_{\mu}\partial_{\nu}h\right) \nonumber\\
&  +\frac{1}{2}\eta_{\mu\nu}\left(  \partial^{2}h-\partial^{\sigma}%
\partial^{\rho}h_{\rho\sigma}\right)  , \label{14}%
\end{align}
where $\partial^{2}\equiv\partial^{\mu}\partial_{\mu}\equiv\square$ and
$h\equiv\eta^{\mu\nu}h_{\mu\nu}.$ To first order in perturbations%
\begin{equation}
\bar{h}^{AB}=\delta_{\mu}^{A}\delta_{\nu}^{B}h^{\mu\nu}+\partial^{A}\chi
^{B}+\partial^{B}\chi^{A}, \label{15}%
\end{equation}
where $h^{\mu\nu}=g^{\mu\nu}-\eta^{\mu\nu}.$ Because $g^{\mu\sigma}%
g_{\sigma\nu}=\delta_{\nu}^{\mu}$ it follows that in linear order $h^{\mu\nu
}=-\eta^{\mu\sigma}\eta^{\nu\rho}h_{\sigma\rho}.$ Now keeping in mind that
capital indices are moved with Minkowski metric $\eta_{AB}$ and replacing them
by Greek indices from (\ref{15}) we find that
\begin{equation}
h_{\mu\nu}=-\bar{h}_{\mu\nu}+\partial_{\mu}\chi_{\nu}+\partial_{\nu}\chi_{\mu
}. \label{16}%
\end{equation}
Substituting (\ref{16}) in (\ref{14}) we find that all $\chi$ terms cancel and
hence $G_{\mu\nu}(h_{\rho\sigma})=-G_{\mu\nu}(\bar{h}_{\rho\sigma})$, that is
the linearized Einstein tensor can be expressed entirely in terms of gauge
invariant variables $\bar{h}_{\rho\sigma}$ as it must be. Then taking into
account that $\bar{h}_{00}=0$ due to constraint (\ref{10}) the linearized
Einstein equations take the form%
\begin{align}
G_{00}\left(  -\bar{h}_{\rho\sigma}\right)   &  =2\lambda+\frac{m^{2}}{4}%
\bar{h},\label{171}\\
\qquad G_{0i}\left(  -\bar{h}_{\rho\sigma}\right)   &  =-\frac{m^{2}}{2}%
\bar{h}_{0i},\label{172}\\
G_{ij}\left(  -\bar{h}_{\rho\sigma}\right)   &  =-\frac{m^{2}}{2}\left(
\bar{h}_{ij}-\frac{1}{2}\eta_{ij}\bar{h}\right)  , \label{173}%
\end{align}
where $G_{\mu\nu}(-\bar{h}_{\rho\sigma})$ are given in (\ref{14}) with
$h_{\rho\sigma}$ replaced by $-\bar{h}_{\rho\sigma}$. It is easy to see that
equations (\ref{11}) and (\ref{12}) follow from the ten equations (\ref{171}),
(\ref{172}), (\ref{173}) as a consequence of Bianchi identities $\partial
^{\mu}G_{\mu\nu}=0.$ Ten equations are enough to determine all ten unknown
variables $\lambda,\bar{h}_{0i}$ and $\bar{h}_{ij}$ (recall that $\bar{h}%
_{00}=0$ due to mimetic constraint). Let us start with $i-j$ equations and
first simplify $G_{ij}(-\bar{h}_{\mu\nu}).$ Using equation (\ref{12}) we find
that%
\begin{equation}
\partial_{i}\partial^{\rho}\bar{h}_{\rho k}+\partial_{k}\partial^{\rho}h_{\rho
i}=\partial_{i}\partial_{k}\bar{h}, \label{18}%
\end{equation}
and
\begin{equation}
\partial^{\sigma}\partial^{\rho}\bar{h}_{\rho\sigma}=\partial^{0}\left(
\partial^{\rho}\bar{h}_{\rho0}\right)  +\partial^{k}\left(  \partial^{\rho
}\bar{h}_{\rho k}\right)  =\frac{1}{2}\partial^{2}\bar{h}+\frac{4\ddot
{\lambda}}{m^{2}}, \label{19}%
\end{equation}
where we have used (\ref{11}) to express $\partial^{\rho}\bar{h}_{\rho0}$ in
terms of $\bar{h}$ and $\lambda$ and dot denotes time derivative. Taking this
into account\ the $i-j$ equations (\ref{173}) become%
\begin{equation}
\partial^{2}\bar{h}_{ij}-\eta_{ij}\left(  \frac{1}{2}\partial^{2}\bar{h}%
-\frac{4\ddot{\lambda}}{m^{2}}\right)  =-m^{2}\left(  \bar{h}_{ij}-\frac{1}%
{2}\eta_{ij}\bar{h}\right)  . \label{20}%
\end{equation}
It immediately follows from (\ref{20}) that the traceless part of spatial
metric components%
\begin{equation}
\bar{h}_{ij}^{T}\equiv\bar{h}_{ij}-\frac{1}{3}\eta_{ij}\bar{h}, \label{22}%
\end{equation}
satisfy the wave equation
\begin{equation}
\left(  \square+m^{2}\right)  \bar{h}_{ij}^{T}=0, \label{23}%
\end{equation}
which describes the massive graviton with five degrees of freedom. The $0-0$
equation (\ref{171}) gives
\begin{equation}
\Delta\bar{h}+\partial^{i}\partial^{j}\bar{h}_{ij}=4\lambda+\frac{m^{2}}%
{2}\bar{h}, \label{24}%
\end{equation}
where $\bigtriangleup=-\partial^{i}\partial_{i}$, when combined with
(\ref{22}) allows us to express $\bar{h}$ just in terms of $\lambda$ and
$\bar{h}_{ij}^{T}$
\begin{equation}
\bar{h}=6\left(  \frac{\partial^{i}\partial^{j}\bar{h}_{ij}^{T}-4\lambda
}{3m^{2}-4\bigtriangleup}\right)  . \label{25}%
\end{equation}
Substituting this expression into the trace of equation (\ref{20}),%
\begin{equation}
\left(  \square+m^{2}\right)  \bar{h}-\frac{24}{m^{2}}\ddot{\lambda}=0,
\label{26}%
\end{equation}
and taking into account that $\bar{h}_{ij}^{T}$ satisfy (\ref{23}) we obtain
the equation which describes mimetic matter%
\begin{equation}
\ddot{\lambda}+\frac{m^{2}}{4}\lambda=0. \label{28}%
\end{equation}
Finally to determine $\bar{h}_{0i}$ we need $0-i$ Einstein equations. To
simplify them we use equation (\ref{11}) to express $\partial_{i}%
\partial^{\rho}\bar{h}_{\rho0}$ in (\ref{14}) in terms of $\bar{h}$ and
$\lambda.$ As a result equation (\ref{172}) takes the form%
\begin{equation}
\Delta\bar{h}_{0i}+\partial_{0}\partial^{k}\bar{h}_{ki}+\partial_{0}%
\partial_{i}\left(  \frac{4}{m^{2}}\lambda-\frac{1}{2}\bar{h}\right)
=m^{2}\bar{h}_{0i}, \label{30}%
\end{equation}
from which, using (\ref{25}) and (\ref{22}), one gets%
\begin{equation}
\bar{h}_{0i}=\frac{\partial_{0}\partial^{k}\bar{h}_{ki}^{T}}{m^{2}-\Delta
}-\frac{\partial_{0}\partial_{i}}{m^{2}-\Delta}\left(  \frac{\partial
^{l}\partial^{m}\bar{h}_{lm}^{T}}{3m^{2}-4\bigtriangleup}\right)
+\frac{16\partial_{0}\partial_{i}\lambda}{m^{2}\left(  3m^{2}-4\bigtriangleup
\right)  }. \label{31}%
\end{equation}
\ 

Thus we have found that massive mimetic gravity describes a massive graviton
characterized by the traceless part of $\bar{h}_{ki}^{T}$ obeying equation
(\ref{23}) and mimetic matter described by $\lambda,$ which satisfies
(\ref{28}). The remaining variables $\bar{h}$ and $\bar{h}_{0i}$ are entirely
expressed in terms of $\bar{h}_{ki}^{T}$ and $\lambda$ (see (\ref{25}) and
(\ref{31})). Mimetic matter is modified and instead of being imitating dust
behaves like particles at rest (with zero momentum) of mass equal to half of
the graviton mass. In the models of massive gravity usually considered with
Fierz-Pauli mass term, corresponding to a combination of $\bar{h}^{2}-\bar
{h}^{AB}\bar{h}_{AB},$ Bianchi identities enforce the vanishing of the
perturbations of the scalar curvature $\delta R=0.$ In turn this leads to vDVZ
discontinuity \cite{vDVZ}, which is resolved only after taking into account
the nonlinear corrections \cite{Vein}. In our case the Bianchi identities
impose conditions (\ref{11}) and (\ref{12}) which are similar to the harmonic
gauge choice, although here these equations are gauge invariant and do not
depend on a coordinate system. Therefore, vDVZ discontinuity is already absent
at the linear level. In other models considered before, the mass term we used
leads to a ghost mode, which is manifested by the dynamics of the $h_{00}$
metric component in the linear theory. In our theory this mode is constrained
and replaced by mimetic matter, which can also explain the observed dark
matter in the universe. This guarantees that mimetic massive gravity is fully
ghost free to all orders \cite{CMlong}. Moreover, in linear order the mimetic
matter is completely decoupled from the graviton.

To consider the massless limit it is convenient to decompose $\bar{h}_{ki}%
^{T}$ in irreducible pieces with respect to the rotation group $SO\left(
3\right)  $%
\begin{equation}
\bar{h}_{ki}^{T}=\left(  \partial_{k}\partial_{i}+\frac{1}{3}\eta_{ki}%
\Delta\right)  S+\partial_{k}V_{i}+\partial_{i}V_{k}+\bar{h}_{ki}^{TT},
\label{32}%
\end{equation}
where $V_{i}$ is transverse $\partial^{i}V_{i}=0$ and $\bar{h}_{ki}^{TT}$ is
not only traceless but also transverse $\partial^{i}\bar{h}_{ki}^{TT}=0.$
Thus, the five degrees of freedom of massive graviton are represented by one
scalar mode $S,$ two vector modes $V_{i}$ and two tensor modes $\bar{h}%
_{ki}^{TT}.$ In \cite{CMlong} we study the dynamics of these modes separately
and by considering quantum fluctuations show that nonlinear corrections for
the scalar and vector modes become important at the energy scale of order
$m^{1/2}.$ At this scale, they get in strongly coupled regime and decouple
from two transverse degrees of freedom of the graviton which become strongly
coupled only at the Planck scale. Thus, at energies above $m^{1/2}$ the
graviton has only two propagating degrees of freedom.

\textbf{{\large {Acknowledgments}}}

The work of A. H. C is supported in part by the National Science Foundation
Grant No. Phys-1518371. The work of V.M. is supported in part by \textit{The
Dark Universe}\ and the Cluster of Excellence EXC 153 \textit{Origin and
Structure of the Universe}. V.M. thanks ENS, where a part of this work was
completed, for hospitality.

\end{document}